\documentclass{article}
\usepackage{graphicx}

\begin{document}
\centerline{\Large \bf The Past, puzzles, and promise of 6-branes }

\centerline{Donald Marolf}
\centerline{Physics Department, Syracuse University, Syracuse, NY 13244},


\begin{abstract}
The fact that both the D6-brane and the orientifold 6-plane have smooth, 
horizon-free descriptions in M-theory makes them especially
useful in understanding certain aspects of brane physics.  We briefly
review how this connection has been used to understand a number of
effects, several of which are associated with
the Hanany-Witten transition.  One particular
outcome is a "confinement mod 2" effect for zero-branes in the background
of a single D8-brane.  We also discuss an interesting puzzle associated
with flux-expulsion from D6-branes in this context.  Finally, we discuss
the promise of using a similar M-theoretic description of the orientifold
6-plane to understand the consistency of stringy negative energy objects
with the 2nd law of black hole thermodynamics.
\end{abstract}
\date{\today}

\section{Introduction}

This outline for this talk arose in an attempt to find a strong
enough unifying theme in my recent work to keep an audience's interest
throughout a 50 minute talk.  Rather to my surprise, such a theme did
exist and, not only did it run through quite a bit of my recent work, but
it continues to run through planned future work as well.  The theme
concerns a certain tool that one can use to uncover certain non-perturbative
effects in brane physics by concentrating on the case of six-branes.
Thus, the above title was born out of the idea that I would review
past work involving six-branes, present some puzzles presently under study
involving six-branes, and
describe the promising future use of six-branes in addressing what may
at first seem like a completely unrelated question.

The feature that makes six-branes unique
in string theory is that they admit smooth, horizon-free 
strong-coupling descriptions in terms of eleven-dimensional supergravity.
In the case of the D6-brane, the lift to M-theory is the Kaluza-Klein
monopole, while for the orientifold 6-plane \cite{Sei1,Sei2,Sen}  it is the Atiyah-Hitchin
manifold \cite{AH}.  These results 
turn out to provide a handle with which to grasp a variety of
non-perturbative effects in brane physics, and we display a selection of
such results below.  

It turns out that several of the results of interest involve the Hanany-Witten
effect \cite{HW}.  For this reason, we begin with a review of the
smooth picture of this effect and then proceed
to discuss what one can do with six-branes.  Our first application
is the construction \cite{GM} of supergravity solutions
describing the Hanany-Witten effect
in which all branes involved are treated as gravitating objects that
affect the bulk spacetime fields (and thus the other branes).  This
construction then leads to a puzzle \cite{Tdual} involving
a certain `flux-expulsion' property of the D6-brane.  

We then turn in a rather different direction to discuss how
D6-branes may be used to derive and understand
a certain `confinement mod 2' effect of D0-branes
in the (symmetric) background created by a unit charged D8-brane \cite{half}.
Finally, we make a further radical change in direction to discuss
the issue of the
consistency of negative tension string-theoretic constructions with
black hole thermodynamics and how the study of 
orientifold 6-planes promises to provide a resolution.

Despite the wide variety of physical questions that will be discussed,
all of these issues
will be studied using the same basic fact that six-branes have
an easily controlled strong coupling description.
While the absence of a gravity/gauge-theory duality \cite{ISMY}
for D6-branes may sometimes make these branes seem less exciting than
their lower dimensional cousins,  I hope that 
the reader is impressed with the variety of issues that
can be raised, addressed, and resolved in the context of six-branes.

\section{The smooth picture of the Hanany-Witten effect}
\label{shn}

This section provides
a brief review of how the Hanany-Witten brane-creation effect \cite{HW}
is described as
a smooth process.  While this discussion has nothing to do with
six-branes specifically, it will set the context and provide background
for two of the sections that follow.
The basic picture follows from general principles,
but one can also find a one-parameter moduli space of exact
BPS solutions describing certain
versions of the process in either the worldvolume theory of a test
brane in the background generated 
by another brane \cite{NOYY,Im,CGS} or in full
supergravity \cite{GM}, meaning that both branes are fully coupled to
bulk fields and can affect each other.  However,  
in this latter case only so-called `near core' solutions are available.  
  What happens in
either the worldvolume theory \cite{wv} or supergravity \cite{3Q}
is that the flux of a gauge
field generated by one brane falling on the second brane generates a third kind
of charge associated with the new brane.
                     
In the supergravity description, this effect follows from the fact
that the `brane-source' charge of the D4-brane (see \cite{3Q}) is not
conserved \cite{GM,3Q}.  This in turn is a straightforward consequence
of the modified Bianchi identity satisfied by the gauge invariant
Ramond-Ramond
four-form field strength $\tilde F_4 = dC_3 + A_1 \wedge H_3$ 
of which the D4-brane is a magnetic
source.  We have the relation
\begin{equation}
\label{mbi}
d\tilde F_4 + F_2 \wedge H_3 = *j^{bs}_{D4},
\end{equation}
where the right hand side is the brane-source current (which vanishes
in the absence of an explicit D4-brane source).  Here, $F_2$ is the
usual IIA Ramond-Ramond two-form field strength and $H_3 = dB_2$ is the
Neveu-Schwarz field strength.  Taking an exterior derivative of
(\ref{mbi}) shows that $d*j^{bs}_{D4}$ does not vanish.  Instead,
a flux of $F_2$ falling on an NS5 brane (where $*j_{NS5}^{bs} \equiv
dH_3 \neq 0$) or a flux of
$H_3$ falling on a D6-brane (where $*j^{bs}_{D6} \equiv dF_2 \neq 0$)
acts as a source or
sink of D4-brane charge.  Some of the subtleties of defining charge and working
with brane-source currents are discussed in \cite{3Q}, but it is enough for
us that this result leads to the Hanany-Witten effect and the associated
creation of a D4-brane as described below.

The diagram below shows various stages in this process
for the case of an NS5-brane moving past a D6-brane to make a D4-brane
\cite{GM}.
Similar results also follow for D$p$ and
D$p'$ branes whenever $p + p' = 8.$, see e.g. \cite{Im,CGS} for a worldvolume
description of the D3/D5 case.
At stage (i) when the NS5-brane is far from the D6-brane,
the center of the
NS5-brane subtends a small angle at the D6-brane and captures only a small
amount of flux from the D6.
As a result, essentially no D4 charge is induced in the region shown
and one has only
a flat NS5-brane.  Then, as the NS5-brane approaches the D6-brane (ii),
it subtends a larger angle and begins to capture some flux, generating
some D4 charge.  This charge corresponds to D4-branes
lying inside the NS5-brane and running outward along this brane
to infinity.
                          
\begin{figure}[h]
\includegraphics{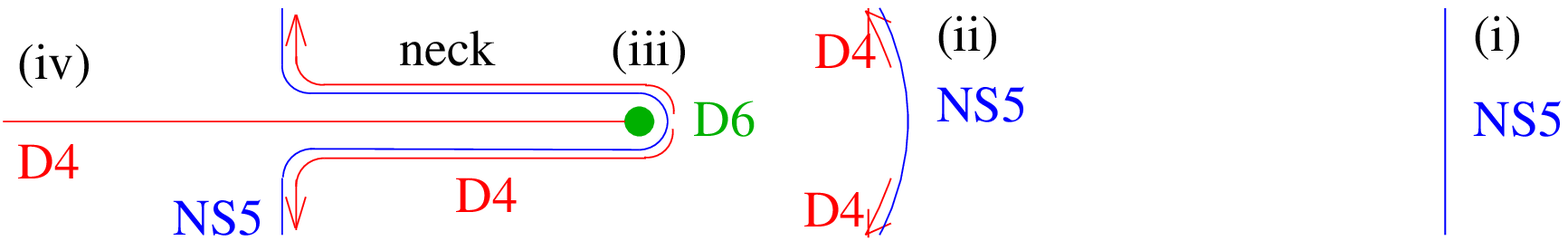}
\caption{Figure 1.  Four stages of the Hanany-Witten process}
\end{figure}

When the NS5-brane is dragged past the D6-brane (iii), all
of the flux from the D6-brane is captured in the part of the NS5-brane close
to the D6-brane.  Capturing one quantum of flux corresponds to the creation
of one quantum of fundamental string charge, so that the thin neck of
NS5-brane approximates a single D4-brane.  However, the NS5-brane
captures flux of the opposite sign in
the region where the neck joins the asymptotically flat part of
the NS5-brane.  The flux captured in this region is half of that
generated by the D6-brane, so that a net one-half quantum of D4 charge
reaches infinity along the NS5-brane.  This last statement is true
in each of the stages (i,ii,iii,iv), though only in stage (iii) are
all of the relevant parts of the NS5-brane visible in figure 1.
In stage (iv), the neck has
narrowed so as to become difficult to resolve and all that remains is a D4-brane
string stretching between an NS5-brane and a D6-brane.    

The above picture seems to follow from general properties of the 
supergravity field equations, but it is important to check them
by studying exact solutions in detail.  This will be particularly clear
in a moment when we discuss the 6-brane `puzzle,' which is an apparent
exception to the above story.  The known exact solutions come in several
forms, the first of which \cite{NOYY} considers a test D2-brane in
the background created by a six-brane.   This case is particularly
tractable using the M-theory description in which we have an M2-brane
in a Kaluza-Klein monopole background.  In this case, any holomorphic
curve represents a BPS configuration of the M2-brane.  By moving the
M2-brane past the monopole, one can watch the formation of a 
string that connects
the monopole to the two-brane.  Test brane solutions were also studied
in \cite{Im,CGS} for the case of a fundamental string stretching between
a D5-brane and a D3-brane, but in this case one must work much harder
to solve the differential equations for the BPS configuration as one does
not have the shortcut of simply looking for holomorphic curves.

\section{Supergravity Solutions and a Puzzle}
\label{puz}

It is interesting, however, to go one step further and to solve
for the full supergravity solutions beyond the test brane approximation; i.e., 
to go to the stage 
in which both branes are actively coupled to the bulk field.  One would
expect that this would show a `back-reaction' of the D2-brane on the
D6-brane.  It turns out that such solutions can in fact be constructed
in what is known as the `near-core limit' using a simple trick
introduced by \cite{ITY,Aki}.  

The key point is that the charge N IIA D6-brane solution lifts to the charge N
Kaluza-Klein monopole solution in M-theory.  In particular, for the unit
charge case the M-theory solution is completely smooth and so is well
approximated by flat space at the center.  As a result, there is a
Kaluza-Klein reduction of flat space that yields the leading approximation
to the D6-brane geometry near the singularity.  This is the `near-core'
D6-brane solution.  The observation is that it is straightforward
to add another brane to this flat space and thereby obtain 
the `near-D6 brane' part of a solution in which the
D6-brane intersects an F1- or D4-brane.  These solutions follow by
simply applying the same
Kaluza-Klein reduction to the M-theory solution describing `an M2- or
M5-brane in flat space;' i.e., to the usual M2- or M5-brane solution. 
This process was begun in \cite{Aki} and completed in \cite{GM}, where
it was shown that an appropriate family of such reductions in fact describes
the near-D6 brane versions of stages (i-iv) in the Hanany-Witten
process for a D2- or NS5-brane being pulled past a D6-brane.  Similarly, 
the multiply charged case can be obtain by first taking an orbifold
quotient and then reducing the result.
   
We refer to the reader to \cite{GM} for the details of these solutions, 
but we mention here an interesting puzzle that one finds after a bit
of study.  As already mentioned, one would expect that constructing
such full supergravity solutions would show the `back-reaction'
of the D2- or NS5-brane on the D6-brane.  Certainly, a brane-charge
argument indicates that, for example, any $\tilde{F}_4$ or $H_3$ flux 
falling on the D6-brane must result in the creation of F1- or D4-branes.
However, one does not see this in the solutions of \cite{GM}.
Instead, there seems to be a `flux-expulsion'
effect associated with D6-branes which is reminiscent of the `superconducting
branes' phenomenon \cite{sup}.   

To begin to understand this effect, 
consider any massless type IIA solution containing
D6-branes.  This of course provides a solution to 11-dimensional
supergravity in which the D6-branes are replaced by the cores of Kaluza-Klein
monopoles.  This solution has a Killing field $\lambda^{11}$ which vanishes
at the core of each monopole.
The natural boundary condition to impose on the D6-branes is that
the corresponding 11-dimensional solutions (or an appropriate
multiple cover in the multiply charged case) be smooth
at these cores.
But now consider the 11-dimensional four-form field  strength
$F_4^{\{11\}}$.  If it is smooth then $F_4^{\{11\}} \cdot
\lambda^{11}$ must vanish when $\lambda^{11}$ does and in particular at
any core.  Since $H_3 = F_4^{\{11\}} \cdot \lambda^{11}$, it follows that
$H_3$ will vanish at any D6-brane.  Note that since
the lowest Fourier mode around the circle will again give some smooth
field, this conclusion also holds in cases
where the 11-dimensional solution does not have an exact translation
symmetry along $\lambda_{11}$ but which can be treated perturbatively.
The same argument also applies to the dual field, so that
$*\widetilde{F}_4=*_{11}F_{4}^{\{ 11 \} }$ should also vanish at a D6-brane.
Here $*_{11}$ is the eleven-dimensional Hodge dual.  This is
the flux that causes D6-branes to produce fundamental strings, so
no fundamental string
charge should be induced on a D6-brane when a D2-brane is dragged past it.
This is also related to a surprising property of a T-dual type IIB solution
\cite{Tdual} involving D5-branes and Kaluza-Klein monopoles.
                                                                        
A similar sort of flux-excluding property was
studied in \cite{sup}.  For the `superconducting'
branes considered in that work, the normal component of some field
strength was forced to vanish on the horizon.  The situation here
is somewhat different, however, as now
the entire field strength $H_3$ or $*\widetilde{F}_4$ must vanish at the brane.
   
Although the above 11-dimensional argument for flux-expulsion meshes
nicely with the unexpected results of \cite{GM} and \cite{Tdual},
certain aspects of this story remain quite puzzling.
For example, flux is clearly
not expelled from the NS5-brane or from a corresponding D2-brane crossing
a D6-brane.  Yet, the D6-brane is connected to these other branes by dualities.
Thus, at least naively it appears that application of supergravity dualities
can transform the solutions of \cite{GM} into ones in which D6-branes
do in fact admit flux from other branes.  Nevertheless, 
finding the mechanism through which this works, or what
alternative resolution string theory provides remains a puzzle to be
solved by further study of D6-branes.

\section{Confinement and Charge quantization}
\label{confine}

We now turn to a related puzzle \cite{half} which, though we will use 6-branes
in its study, is most easily stated in terms of D0- and D8-branes.
Consider for example a system with a single D0-brane and a single D8-brane and
suppose that the boundary conditions are such that the ten-form Ramond-Ramond
gauge field takes the symmetric values $\pm 1/2$ of the fundamental quantum
on either side of the D8-brane
domain wall.  Then, the brane-source charge arguments above (or, 
equivalently the arguments of \cite{PS,BGL} in type IIA supergravity 
\cite{Romans} and the arguments of
\cite{Lif,BDG,DFK,K,HoWu,dA,OSZ} in related contexts)  
lead to the conclusion
that exactly 1/2 of a fundamental string must end on the D0-brane.  While this
seems to be at odds with charge quantization, several possible resolutions
immediately present themselves.  One possibility is that the half-string is
a mere artifact of some accounting scheme
(see, e.g. \cite{BDS,Taylor,Mor,3Q,SS})
and that it
is not in fact in conflict with charge quantization.  Another
possibility is that such symmetric boundary conditions for
D8-branes are not actually allowed,
and that the Ramond-Ramond gauge ten-form field strength $F_{10}$
must take integer values.
A final possibility is that $F_{10}$ is allowed to take half-integer
values but that, in such backgrounds, D0-brane charge is allowed to occur
only in multiples of 2.  We will conclude that this final scenario is
in fact correct by considering the T-dual D2/D6 system and again using
the description of D6-branes as Kaluza-Klein monopoles in M-theory.

The same question of course arises for the D2/D6 case.  It is useful to first
establish notation and we recall
that, supposing the D6-brane is oriented along the $x_0,x_1,x_2...x_6$, 
the unit charged D6-brane solution takes the form

\begin{eqnarray}
\label{10Dsol}
ds^2_{string} &=&  V^{-1/2} dx_\parallel^2 + V^{1/2} dx_\perp^2, \cr
e^{2\phi} &=& V^{-3/2}, \cr
A_1 &=&  \frac{1}{2} (1 - \cos \theta) d \psi, \cr
F_2 &=& \frac{1}{2} \sin \theta  d \theta \wedge d\psi
\end{eqnarray}
where we have introduced $dx_{\parallel}^2 = -dx_0^2 + dx_1^2
+ dx_2^2 + dx_3^2 + dx_4^2 + dx_5^2 + dx_6^2$ and
$dx_\perp^2 = dx_7^2 + dx_8^2 + dx_9^2$, along with
$V = 1 + \frac{1}{2r}$, $r =x_7^2 + x_8^2 + x_9^2$,
$\theta = \cos^{-1} \left( \frac{-x_9}{r} \right)$,
and $\psi = \tan^{-1}\left( \frac{x_8}{x_7} \right)$.
Here, to simplify the formulas we have set the radius $R_{10}$
of the M-theory circle to one.

If the D2-brane
is extended in two directions (say, $x_7,x_8$) 
orthogonal to the D6-brane, then it will
capture half of the flux from the D6-brane and must therefore have
half of a fundamental string ending on the D2.  In \cite{half}, this question
was studied using the method of \cite{NOYY}; i.e., by considering
test M2-branes in the Kaluza-Klein monopole background.  There certainly do exist
configurations in which the D2-brane is extended orthogonally to the D6-brane, and for
these cases the issue is merely one of proper accounting.  In particular, while
there is indeed 1/2 unit of fundamental string `brane source' charge in this
system (in particular, this charge can be shown to flow 
along the D2-brane world-volume to infinity), brane-source
charge is not in general quantized (see \cite{3Q}).  Instead, the
measure of charge that is 
quantized is known as the `Page charge.'   
While the value of this charge is not gauge invariant, its value for this
configuration is an
integer in any gauge \cite{half}.  In particular, in simple
gauges one finds either zero or one units of fundamental string charge.

However, one can show \cite{half} that the distinction between brane-source and Page charge is 
important only for the case that the fundamental string 
charge runs to infinity along
the worldvolume of the D2-brane\footnote{In the usual worldvolume description, one would represent
this as a non-zero flux to infinity of the D2-brane gauge field}.  Note, however, that the flux of
fundamental string charge or world-volume gauge field to infinity will obstruct any attempts to
compactify this solution in the directions along the D2-brane.  Because the flux is only outward, no
consistent 
identifications can be imposed on solutions with such a flux.  From the worldvolume perspective, 
this is just the familiar statement that the total charge coupled to the gauge field must vanish
on a compact worldvolume.  As a result, such
configurations cannot be compactified and therefore are not in fact related by T-duality to the
D0/D8 case.    

Thus, the issue remains.  However, we have learned that we must focus on the case in which
no fundamental string brane-source charge flows along the D2-brane to 
infinity.  Nonetheless, the D2-brane will necessarily intercept some flux from the D6-brane and it is clear that
some fundamental string charge must somehow flow off of the D2-brane.  The only
remaining possibility is that this charge will in fact flow to the D6-brane itself.  Since
the fundamental string in this context is nothing but a deformation of the D2-brane worldsheet, this
means that we must 
consider solutions where the D2-brane actually intersects the D6-brane.
It is here that the M-theory context is particularly useful, as what appears to be a singular
intersection in the IIA description becomes merely the smooth passage of an M2-brane through
the core of a Kaluza-Klein monopole.  In particular, while the D6-brane singularity would
prevent one from determining the true structure of the intersection from the IIA perspective, from the 
11-dimensional perspective it is clear that valid D2/D6 intersections are exactly those for which the
M2-branes remain smooth at the Kaluza-Klein monopole core.

The general holomorphic such intersection is analyzed in \cite{half}. However, for simplicity we examine here
only the special case in which the 
D2-brane is placed on the surface $x_9=0$.  To describe the
11-dimensional description of this surface, we must first
establish our conventions for the Kaluza-Klein monopole in 11-dimensions.
A useful form of this metric is
\cite{NOYY,Yosh,BG}

\begin{equation}
ds^2 = - dx_\parallel^2 + V dv d\overline v + V^{-1} \left|
\frac{dw}{w} - f dv \right|^2,
\end{equation}
where
\begin{equation}
f = \frac{x_9 + r}{2vr},
\end{equation}
correcting a small typographic error in \cite{NOYY,BG}.
The complex coordinates
$v$ and $w$ define one of the complex
structures on the Euclidean Taub-Nut space.  They are related to the
ten-dimensional coordinates through
\begin{eqnarray}
v &=& x_7 + i x_8 \cr
w &=& e^{-(x_9+ix_{10})} \left( - x_9 + \sqrt{x_9^2 + |v|^2} \right)^{1/2},
\end{eqnarray}
and Kaluza-Klein reduction takes place along the Killing  field
$\partial_{x_{10}}$.
Such coordinates are smooth so long as $v \neq 0$ or $x_9 < 0$.
Note that $x_{10}$ ranges over $[0,2\pi]$ consistent with our
setting $R_{10}=1$.  In addition, a careful check will show that this space 
has a $Z_2$ symmetry of the form  $(w,v) \rightarrow (\frac{v}{w}, v)$.

A holomorphic curve that reduces to the surface $x_9=0$
surface can be found by noticing that
the symmetry $(w,v) \rightarrow (\frac{v}{w},v)$ changes the sign
of $x_9$, so that any surface which is invariant under this symmetry
must lie at $x_9=0$.  The surface
$w^2 = v$ is invariant in this way, and careful investigation
\cite{half} shows that it remains smooth at the origin ($v=w=0$).  
However, because it contains $w^2$,
upon dimensional reduction we find {\it two} D2-branes lying at $x_9=0$.
The corresponding sheets of the M2-brane lie at $x_{10} = \psi$ and
$x_{10} = \psi + \pi.$
Deforming this surface to move the asymptotic parts of both D2 branes
to large $|x_9|$ therefore produces a string-like piece of D2-brane
connected to the D6-brane and carrying a full unit of fundamental string charge and tension.

Note that
the two D2-branes cannot be separated from one another as, when
the angle $\psi$ increases by $2\pi$, we must move from one brane
to the other.  
It turns out (see \cite{half}) that in fact all D2-brane configurations
that can be compactified have a similar structure, with a pair of D2-branes forming a Riemann surface
whose a branch point lies at the location of the D6-brane.  As a result, the same behavior should be found in
the T-dual D0/D8 system.  That is, in the background of a unit charged D8-brane, D0-branes
should only occur in pairs and these pairs must be connected to the D8-brane by (integer) fundamental strings.
We refer to this effect as ``confinement of the D0-branes in pairs.''

It is interesting to remark that a 
${\bf Z}_2$ quotient of the above solution leads
to the charge 2 Kaluza-Klein monopole and the charge 2 D6-brane.
Such a ${\bf Z}_2$ quotient identifies the two sheets of the D2-brane, leading
to a single D2-brane in the charge 2 D6-brane background.  
Thus, the confinement effect disappears in the charge 2 D8-brane background.

\section{The future promise of the six-brane}
\label{orient}

In this final section, we give a preview of other results one may hope to obtain from a study of
six-branes.  This latter study is of quite a different nature, and 
involves the relationship between negative energy objects and the generalized second law of thermodynamics.
Let us first state the issues involved, and we will then fore-shadow briefly why
 a study of six-branes, this
time of the orientifold variety, should allow us to understand the situation.

The following discussion is motived by the recent appearances of negative energy objects in 
various large extra dimension scenarios (see, e.g. \cite{Rubakov:1983bb}-\cite{Corley:2001rt}).
It is, of course, important that any extra-dimensional scenario be stable, and negative
energies can be problematic.   Placing a negative tension brane
at a orbifold fixed plane has been shown to remove perturbative dynamical
instabilities.  Nevertheless, one still worries about the second law
of thermodynamics, particularly as generalized to include the entropy of 
black holes.  To see the point, consider a process in which a bit
of a negative tension brane is lowered into a black hole (or, equivalently, in which a black hole
is thrown at a negative tension brane).  Heuristically, one expects that as negative
energy matter flows into the black hole, the black hole will shrink and reduce in entropy.
In the semi-classical limit, the entropy of a black hole should dominate over all other
forms of entropy and one expects a reduction in black hole area to
lead to a violation of the generalized second law of thermodynamics. 

In \cite{NegT}, exact constructions were given of spacetimes representing collisions in 2+1 dimensional 
gravity between BTZ black holes and 1+1 dimensional negative tension branes at orbifolds.  
The results are interesting in that one
can prove that if the system were to settle
down to some equilibrium state representing a black hole attached to the brane, then a
violation of the second law would necessarily result.  However, in this case such clear violations
are avoided by the onset of a catastrophe.  Instead of settling down to some equilibrium state, 
a new spacelike singularity forms that extends far outside of what one would naively have called the
black hole, reaching out to engulf the entire brane and, in a certain sense, the entire universe.  

One is led to wonder whether this behavior is somehow typical or whether other sorts of branes
are better behaved.  In particular, a priori this might be an artifact of our low-dimensional
setting, and a higher dimensional study is currently in progress in collaboration with
Joel Rozowsky, Pedro Silva, and Mark Trodden.  However, one suspects that the above
catastrophe is somehow related to the more basic issue involving black hole thermodynamics.
It seems likely that negative tension objects should be allowed only in the case that they have
some property that guarantees compatibility with black hole thermodynamics.

This is where we come to the connection with six-branes.  We begin by recalling that
certain negative tension orientifolds do in fact arise in string theory 
(see, e.g., \cite{OGS,Hanany:2000fq} for recent reviews). String theoretic
calculations do not indicate any instabilities of such objects, and given the history
of string theory one might well suspect that such objects could teach us something new
about gravitational physics.  Thus, one expects that such orientifolds are the best
candidates for negative tension objects that may be compatible with black hole thermodynamics.

As with the puzzles discussed above, it appears that the six-plane case can shed further
light on the issues.  It turns out that (at strong coupling) there is an M-theoretic
understanding of the orientifold six-plane \cite{Sei1,Sei2,Sen}
in terms of the Atiyah-Hitchin manifold \cite{AH}.
The resulting 11-dimensional spacetime is in fact a smooth Ricci-flat (hyperk\"ahler) manifold
whose structure near infinity resembles a $Z_2$ quotient of a negative tension Kaluza-Klein
monopole.  As a result, one may describe
the collision of a black hole with such an orientifold as a problem
in pure {\it vacuum} (i.e., source-free)  
Einstein-Hilbert gravity in eleven dimensions.  
Under such conditions,
the Raychaudhuri equation leads in the usual way \cite{HE}
 to the conclusion that the total horizon area must increase during the
collision.  
Our violations of the generalized second law of thermodynamics will not
arise in this context.  As the various stringy negative tension orientifolds are
related by T-duality, one expects that the other orientifolds of string theory
also have properties such that the second law of thermodynamics
is upheld in collisions with black holes.

Since we have seen that the orbifold boundary condition itself
is not sufficient to make negative tension compatible with black hole
thermodynamics, it would be interesting to understand in more detail just what
properties of these orientifolds enforce the second law.  One suspects that
one need only probe the details of the Atiyah-Hitchin manifold for the answers.
A study of this form is currently in progress in collaboration with Simon Ross.
We hope to report the results of this work soon.
                                         
\bigskip

{\bf Acknowledgments}

\medskip

In addition to the various acknowledgments in the original papers, 
I would like to thank  
Andr\'es Gomberoff, Simon Ross, and Mark Trodden for delightful collaborations
on various parts of the work reported here.  
This work was supported in part by
NSF grants PHY97-22362 and PHY00-98747 to Syracuse University,
the Alfred P. Sloan foundation, and by funds from Syracuse
University.


\begin{thebibliography}{99}

\bibitem{Sei1}
N.~Seiberg,
Phys.\ Lett.\ B {\bf 384}, 81 (1996)
[hep-th/9606017].

\bibitem{Sei2}
N.~Seiberg and E.~Witten,
hep-th/9607163.

\bibitem{Sen}
A.~Sen,
JHEP{\bf 9709}, 001 (1997)
[hep-th/9707123].

\bibitem{AH} M. Atiyah and N. Hitchin, {\it The Geometry and Dynamics of Magnetic
Monopoles}, (Princeton, 1988). 

\bibitem{HW} A. Hanany and E. Witten, ``Type IIB
Superstrings, BPS Monopoles, and Three-Dimensional
Gauge Dynamics,'' Nucl.\ Phys.\  {\bf B492}, 152 (1997), hep-th/9611230.

\bibitem{ISMY} 
N.~Itzhaki, J.~M.~Maldacena, J.~Sonnenschein and S.~Yankielowicz,
Phys.\ Rev.\ D {\bf 58}, 046004 (1998)
[hep-th/9802042].

\bibitem{NOYY} T.~Nakatsu, K.~Ohta, T.~Yokono and Y.~Yoshida,
Mod.\ Phys.\ Lett.\ A {\bf 13}, 293 (1998)
hep-th/9711117.

\bibitem{Im}Y. Imamura, Nucl.\ Phys.\ {\bf B537} (1999) 184-202.

\bibitem{CGS}C. Callan, A. Guijosa, and K. Savvidy,
``Baryons and string creation from the fivebrane worldvolume action,''
Nucl.\ Phys.\
{\bf B547} (1999) 127-142, hep-th/9810092.

\bibitem{GM} A. Gomberoff and D. Marolf,
``Brane transmutation in supergravity,''
JHEP {\bf 0002}, 021 (2000),
hep-th/9912184.

\bibitem{wv}  U. H. Danielsson, G. Ferretti, and I. R. Klebanov,
``Creation of Fundamental Strings by Crossing D-branes,''
Phys.\ Rev.\ Lett. 79 (1997) 1984-1987, hep-th/9705084.

\bibitem{3Q} D.~Marolf, ``Chern-Simons terms and the three notions of charge,''
hep-th/0006117.

\bibitem{ITY} N. Itzhaki, A. Tseytlin, and S. Yankielowicz, {``Supergravity
Solutions for branes localized within branes''} {\it Phys. Lett.}
{\bf B432} 298-304 (1998), hep-th/9803103.

\bibitem{Aki} A. Hashimoto, {``Supergravity solutions for localized
intersections of branes,''} {\it JHEP} {\bf 9901} 019 (1999),
hep-th/9812159.          

\bibitem{sup} A.~Chamblin, R.~Emparan and G.~W.~Gibbons,
``Superconducting p-branes and extremal black holes,''
Phys.\ Rev.\ {\bf D 58}, 084009 (1998)
[hep-th/9806017].

\bibitem{Tdual}
D.~Marolf,
JHEP {\bf 0106}, 036 (2001)
[hep-th/0103098].

\bibitem{half} D.~Marolf,
``Half-branes, singular brane intersections, and Kaluza-Klein reduction,''
JHEP {\bf 0009}, 024 (2000),
hep-th/0007171.

\bibitem{PS} J. Polchinski and A. Strominger, {``New Vacuua for Type
II String Theory,''}
Phys.\ Lett.\  {\bf B388}, 736 (1996), hep-th/9510227.

\bibitem{BGL} O. Bergman, M. R. Gaberdiel, and G. Lifschytz,
{\it ``Branes, Orientifolds, and the Creation of Elementary
Strings,''}
Nucl.\ Phys.\  {\bf B509}, 194 (1998), hep-th/9705130.

\bibitem{Romans} L. Romans, {\it Phys. Lett.} {\bf B169} 374 (1986).
 
\bibitem{Lif} G. Lifschytz, {\it Comparing D-branes to Black Branes},
Phys.\ Lett.\  {\bf B388}, 720 (1996),
hep-th/9604156.

\bibitem{BDG} C. P. Bachas, M. R. Douglas, and M. B. Green,
JHEP {\bf 9707}, 002 (1997),
{``Anomalous Creation of Branes,''} hep-th/9705074.

\bibitem{DFK} U. Danielsson, G. Ferretti, and I. R. Klebanov,
{``Creation of Fundamental
Strings by Crossing D-branes,''}
Phys.\ Rev.\ Lett.\  {\bf 79}, 1984 (1997),
hep-th/9705084.

\bibitem{K} I. R. Klebanov, {``D-branes and the Creation of Strings,''}
Nucl.\ Phys.\ Proc.\ Suppl.\  {\bf 68}, 140 (1998),
hep-th/97091160.

\bibitem{HoWu} P.~Ho and Y.~Wu,
``Brane creation in M(atrix) theory,''
Phys.\ Lett.\ B {\bf 420}, 43 (1998),
hep-th/9708137.

\bibitem{dA} S.~P.~de Alwis,
``A note on brane creation,''
Phys.\ Lett.\ B {\bf 413}, 49 (1997),
hep-th/9706142.


\bibitem{OSZ} N.~Ohta, T.~Shimizu and J.~Zhou,
``Creation of fundamental string in M(atrix) theory,''
Phys.\ Rev.\ D {\bf 57}, 2040 (1998),
hep-th/9710218.                

\bibitem{BDS} C. Bachas, M. Douglas, and C. Schweigert, {``Flux
Stabilization of D-branes,''} JHEP {\bf 0005}, 048 (2000),
 hep-th/0003037.

\bibitem{Taylor} W. Taylor, {\it ``D2-branes in B fields''},
hep-th/0004141.

\bibitem{Mor}
A.~Alekseev, A.~Mironov and A.~Morozov,
``On B-independence of RR charges,''
hep-th/0005244.

\bibitem{SS} S. Stanciu, ``A Note on D0-branes in Group Manifolds: Flux
Quantization and D0-charge,'' hep-th/0006145. 

\bibitem{Yosh} Y. Yoshia, {``Geometrical Analysis of Brane
Creation via $M$-theory,''}
Mod.\ Phys.\ Lett.\  {\bf A13}, 293 (1998), hep-th/9711177.

\bibitem{BG} C. P. Bachas, M. B. Green, {``A Classical Manifestation of the
Pauli Exclusion Principle,''} JHEP {\bf 9801}, 015 (1998), hep-th/9712187. 


\bibitem{Rubakov:1983bb}
V.~A.~Rubakov and M.~E.~Shaposhnikov,
Phys.\ Lett.\ {\bf B125}, 136 (1983).

\bibitem{Rubakov:1983bz}
V.~A.~Rubakov and M.~E.~Shaposhnikov,                                          
 Problem,''
Phys.\ Lett.\ {\bf B125}, 139 (1983).

\bibitem{akama}
K. Akama, "Pregeometry" in Lecture Notes in Physics, 176, Gauge Theory and 
Gravitation, Proceedings, Nara, 1982, (Springer-Verlag), 
edited by K. Kikkawa, N. Nakanishi
and H.
Nariai, 267-271

\bibitem{Arkani-Hamed:1998rs}
N.~Arkani-Hamed, S.~Dimopoulos and G.~Dvali,
Phys.\ Lett.\ {\bf B429}, 263 (1998)
[hep-ph/9803315].

\bibitem{Arkani-Hamed:1999nn}
N.~Arkani-Hamed, S.~Dimopoulos and G.~Dvali,
gravity,''
Phys.\ Rev.\ {\bf D 59}, 086004 (1999)
[hep-ph/9807344].

\bibitem{Antoniadis:1998ig}
I.~Antoniadis, N.~Arkani-Hamed, S.~Dimopoulos and G.~Dvali,
Phys.\ Lett.\ {\bf B436}, 257 (1998)
[hep-ph/9804398].
                                         

\bibitem{Randall:1999ee}
L.~Randall and R.~Sundrum,
Phys.\ Rev.\ Lett.\ {\bf 83}, 3370 (1999)
[hep-ph/9905221].

\bibitem{Randall:1999vf}
L.~Randall and R.~Sundrum,
Phys.\ Rev.\ Lett.\ {\bf 83}, 4690 (1999)
[hep-th/9906064].

\bibitem{Arkani-Hamed:2000hk}
N.~Arkani-Hamed, S.~Dimopoulos, G.~Dvali and N.~Kaloper,
Phys.\ Rev.\ Lett.\ {\bf 84}, 586 (2000)
[hep-th/9907209].

\bibitem{Lykken:2000nb}
J.~Lykken and L.~Randall,
JHEP{\bf 0006}, 014 (2000)
[hep-th/9908076].

\bibitem{Sundrum:1999ns}
R.~Sundrum,
Phys.\ Rev.\ {\bf D 59}, 085010 (1999)
[hep-ph/9807348].

\bibitem{Arkani-Hamed:1998kx}
N.~Arkani-Hamed, S.~Dimopoulos and J.~March-Russell,
hep-th/9809124.


\bibitem{Corley:2001rt}
S.~Corley and D.~A.~Lowe,
hep-ph/0101021.

\bibitem{NegT}
D.~Marolf and M.~Trodden,
Phys.\ Rev.\ D {\bf 64}, 065019 (2001)
[hep-th/0102135].

\bibitem{OGS}
O.~Bergman, E.~Gimon and S.~Sugimoto,
JHEP {\bf 0105}, 047 (2001)
[hep-th/0103183].

\bibitem{Hanany:2000fq}
A.~Hanany and B.~Kol,
JHEP{\bf 0006}, 013 (2000)
[hep-th/0003025].

                   
\bibitem{HE} S. W. Hawking and G. F. R. Ellis, {\it The large scale structure
of space-time}, (Cambridge, 1973).

\end{thebibliography}
\end{document}